\documentclass{elsart}

\usepackage[authoryear]{natbib}
\usepackage{graphicx}

\begin{document}
\begin{frontmatter}
\title{Fine tracking system for balloon-borne telescopes}

\author[mp]{M. Ricci\corauthref{cor}},
\corauth[cor]{Corresponding author.}
\ead{massimo.rcc@libero.it}
\author[mp]{F. Pedichini},
\ead{fernando.pedichini@oa-roma.inaf.it}
\author[mp]{D.Lorenzetti},
\ead{dario.lorenzetti@oa-roma.inaf.it}

\address[mp]{INAF - Osservatorio Astronomico di Roma - Via Frascati, 33 -
00040 Monte Porzio Catone (Italy)}

\begin{abstract}

We present the results of a study along with a first
prototype of a high precision system ($\le$ 1 $arcsec$) for
pointing and tracking light (near-infrared) telescopes on board
stratospheric balloons. Such a system is essentially composed by a
{\it star sensor} and by a {\it star tracker}, able to recognize
the field and to adequately track the telescope, respectively. We
present the software aimed at processing the star sensor image and
the predictive algorithm that allows the fine tracking of the
source at a sub-pixel level. The laboratory tests of
the system are described and its performance is analyzed. We demonstrate
how such a device, when used at the focal plane of enough large telescopes
(2-4m, F/10), is capable to provide (sub-)$arcsec$ diffraction limited images
in the near infrared bands.
\end{abstract}


\end{frontmatter}

\section{Introduction}

The excellent conditions present in the high stratosphere (25-35
km) especially for near-infrared (NIR) observations allow
balloon-borne telescopes approaching space telescopes
performances. The stratosphere is very dry, which minimizes
atmospheric opacity, an advantage magnified by the low pressure
which reduces line broadening. Moreover, the ambient temperature
is also low (T=220-240 K), which minimizes thermal emission from
the telescope optics and reduces atmospheric radiance. Finally,
low turbulence reduces seeing to negligible effects.

On the other hand, observations from aircraft suffer from image
degradation due to air turbulence and vibration. Observations from
space require the development of new technological aspects and are
intrinsically very expensive. Balloon flights, although offer an
observing time limited by the flight duration, provide conditions
close approaching to space ones, but at much more reasonable
costs. Well known problems related to the realization of a large
balloon-borne NIR telescope mainly concern: ({\it i}) the use of
lightweight mirrors to keep the total weight of the experiment
below 1000-1500 kg; ({\it ii}) the realization of long duration
balloons which can remain at 30 km of altitude for several weeks;
({\it iii}) accurate attitude control to keep pointing stability
at the level of 0.05 arcsec; ({\it iv}) recovery of the payload
and telescope for next flights.
 New positive
results are being obtained in some of the technological aspects
mentioned above. SiC mirrors up to 1m have been realized with
optical quality and larger mirrors are coming soon (e.g. Webb
2007; Kaneda et al. 2007). Their surface weight is of the order of
20 kg/m$^2$ implying for a 4m diameter mirror a total weight of about
only 400 kg. Long duration flights up to 40 days have
been recently obtained in cosmic ray experiments in Antarctica
(CREAM experiment - Beatty et al. 2003). Development of AO
facility for large ground-based telescopes allows the realization
of innovative lightweight optical systems able also to keep an
excellent pointing stability. Finally, airbag technology for the
soft landing of experiments on the surface of external planets
could allow the realization of a safe recovery of the telescope
(eg. http://www.lockheedmartin.com/). Remarkably, many of these
issues were afforded (and successfully solved) more than 40 yrs
ago (Stratoscope II Balloon-Borne Telescope - Danielson et al.
1972; Wieder 1969; McCarthy 1969).

In the present paper we present our preliminary results obtained
in the framework of the attitude control to keep pointing
stability within acceptable levels. This paper is structured as
follows: in Sect.2 the modalities of the gondola oscillations are
described, while in Sect.3 we present both software procedures and
hardware components we have developed to remotely recognize the
observed field. The fine tracking system is fully described in
Sect.4, along with the results of our laboratory tests. Our concluding
remarks are given in Sect.5.

\section{Stratospheric balloon gondola oscillations}

The correct pointing of a telescope on board a stratospheric
balloon is mainly hampered by the payload pendular motion which is
essentially due to the residual atmospheric turbulence still
present at a stratospheric altitude. The most evident effect of
such turbulence is to provoke balloon rotation and, to a lesser
extent, to make the roll and pitch angles oscillating.

Usually, azimuth rotation is reduced by some decoupling devices
located in between the gondola and its suspension cable: they aim
at both nulling the torsion of this latter and keeping the payload
in a state of rest. Residual oscillations typical of this state
are characterized by a less than 1 {\it arcmin} amplitude and by a
sub-Hz frequency.

Roll and pitch angles present oscillations of small amplitude at
low frequency, as well; typical values of the balloon and gondola
frequencies are 0.1 and 1 Hz, respectively (Fixsen et al. 1996):
our fine tracking system aims at correcting such residual
oscillations. It is based on three different control levels:

\begin{itemize}
\item[1 -] A dedicated Inertial Measurement Unit (IMU), based on
gyroscopes, computes instant angular velocities and provides the
telescope servo-motors with suitable correction signals to keep
its pointing well within 1 arcmin.
\item[2 -] A {\it star sensor} camera computes the absolute
pointing with enough accuracy ($<$ 1 arcmin); gives the signal for
IMU long-term drift compensation, and improves the telescope
coarse pointing until the target field is reached.
\item[3 -] Once on target, a fine optical tracking system locks on
one (or more) reference star(s) with a 0.1 pixel accuracy by using
a predictive algorithm. Field rotation of the focal plane instrumentation
is supposed to be provided by an usual mechanical system (if needed,
gyroscope assisted) able.
\end{itemize}

\section{The star sensor}

The star sensor gives the equatorial coordinates of the observed
field center, in order to recognize the sky area pointed by the
telescope. It uses a CMOS camera whose field of view (about
50$^{\circ}$) is enough to allow us observing, at least, three
bright ($<$ 3 mag) stars. After an image analysis, a built-in
software procedure provides the astrometric solution in equatorial
coordinates of the field center.

\subsection{Recognition of the stellar fields}

\begin{figure}\hspace{3cm}
\includegraphics[width=7cm]{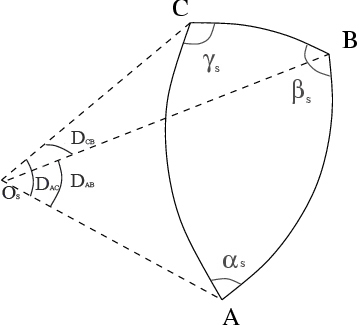}
\caption{\label{ABC_stars} Sketch of three generic stars named A,B, and C on the celestial sphere. The
corresponding angles are $\alpha_s$, $\beta_s$, and $\gamma_s$, while their relative angular distances
are D$_{AB}$, D$_{AC}$, and D$_{CB}$. These latter are needed for the stars identification and for the
comparison with the stellar catalog. O$_S$ indicates the center of the celestial sphere.}
\end{figure}

For recognizing the stellar field the following procedure is
adopted. From any image, the coordinates (in pixels) of the bright
stars are derived, and the angular distances between them are
computed (see Figure~\ref{ABC_stars}) as a function of the
parameters associated to the optical system. The software
procedure is able to identify stars in any sky image by comparing
their angular distances with those ones given by a stellar
catalog, assuming that the obtained images are gnomonic projection
of the celestial sphere. Such a comparison is done through the
relationship:

\begin{equation}
{\cos(ang. distance) = \sin(\delta_1) \sin(\delta_2) +
\cos(\delta_1) \cos(\delta_2) \cos(\alpha_1 - \alpha_2)}
\end{equation}

where $\alpha_1$, $\alpha_2$, $\delta_1$,, $\delta_2$ are the
Right Ascension (RA) and the Declination (Dec) of a couple of
stars. In this way, the three stars selected for providing the
astrometric solution are identified. The center field RA and Dec
are found by means of a numerical and analytical solution,
respectively.

\vspace{3cm}

\subsection{Test with real images}

\begin{figure}[t!]
\includegraphics[width=14cm]{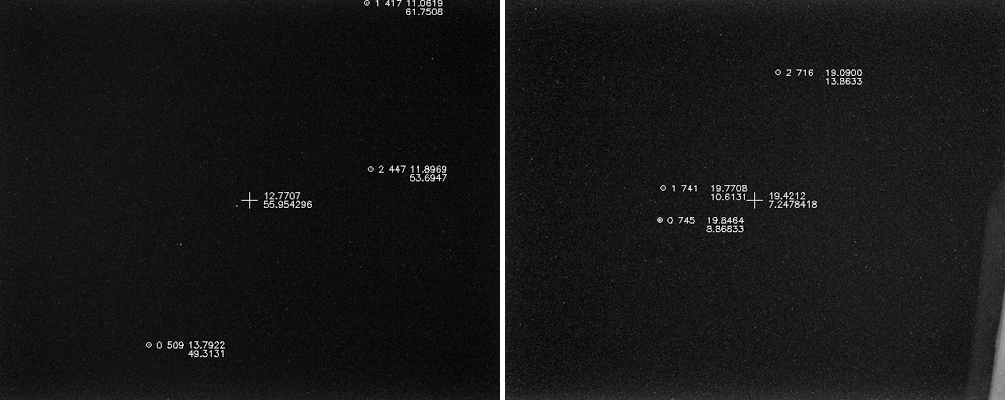}
\caption{\label{sky} Images of the Ursa Mayor constellation (left
panel) and Aquila (right panel) obtained with our camera (see
text). The stars selected to obtain the astrometric solution are
indicated with both a progressive number (0, 1, 2,..) assigned by
our software and the corresponding catalog (J2000 NEW FK5)
identification. RA (in hours) and Dec (in degrees) complete the
label of any recognized star. A cross indicates the field center
(in equatorial coordinates).}
\end{figure}

A star sensor prototype has been assembled by using a 1.3
Megapixel CMOS detector (e2v EOS-AN012) with a pixel size of
5.3$\times$5.3 $\mu$m, equipped with a 12mm F/3.0 objective
(Sunsex DSL901C). The field of view corresponds to $\approx$
50$^{\circ}$ with an angular resolution of 3$^{\prime}$/pixel; our
calibration tests did not evidenced any significant optical
distortion. In Figure~\ref{sky} the images obtained for two
constellations are depicted: sky test results are fully compatible
with the camera resolution and indicate an absolute pointing error
$<$ 1 pixel (i.e. $<$ 3$^{\prime}$).

\section{The star tracker}

The star tracker performs a real time correction of the image
smearing due to the payload oscillations. At variance with other
tracking systems based on PID (Proportional Integral Derivative)
correction algorithms, the one described here presents a novelty,
namely it is conceptually based on predicting the position that
the reference source centroid is expected to have. This is
accomplished through the mathematical analysis of a stationary
time series.

\subsection{Predicting the balloon trajectory}

For predicting the trajectory, an auto-regressive model is used: it
allows us to write the predicted value x$_t$ of a time series
X = [x$_0$, x$_1$, x$_2$, ..., x$_{t-1}$] as a linear combination
of p values, already measured, to which a random error z$_t$ is
added:

\begin{equation}
{x_t = a_1x_{t-1} + a_2x_{t-2} + ... + a_px_{t-p} + z_t}
\end{equation}

It works in a way similar to a model of linear regression, in which,
however, x$_t$ is not defined by independent variables, but by its
already measured values. The coefficients a$_1$, a$_2$, ..., a$_p$ are
computed in order to minimize the uncorrelated term z$_t$.
Noticeably, the order p of the auto-regressive process that best
represents a given time series, is hard to be derived: a
viable approach consists in increasing progressively the order of
the process until the sum of the residuals reaches a required
value (Chatfield, 1995). In our case, the time series
representing an oscillating gondola is well described by an
auto-regressive process of order 4.

\subsection{Testing the predictive software}

The experimental test bench of the fine tracking predictive
software simply consists in a 1.3 Mpixels CMOS camera MAGZERO
MZ-5m, with a lens of 100mm focal length (corresponding to a plate
scale of 10 arcsec/pixel). Moreover, to simulate the payload
pendular motion, the camera has been stiffly mounted over an
oscillating optical bench (at $\sim$ 2 Hz). The reference source
is a light spot from an optical fiber.

About 100 images have been taken (each 10 msec integrated,
repeated every 50 msec). The predictive software analyzes these
images and provides the predicted position to be compared with the
real one. Thus, the goodness of our prediction can be evaluated.

\begin{figure}[t!]
\includegraphics[width=14cm]{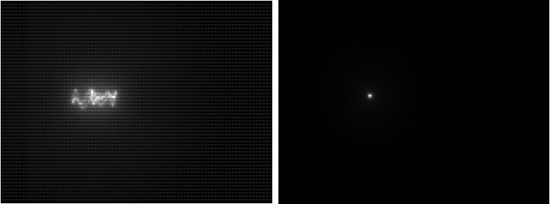}
\caption{\label{sum} {\it Left Panel}: co-adding of 100 subsequent
images, each taken every 50 msec, and 10msec highlights the real
trajectory of the oscillating bench. {\it Right Panel}:
re-centered co-adding operated by the predictive software.
Evidently, the effects of the bench oscillations are negligible here.}
\end{figure}

\begin{figure}[t!]
\includegraphics[width=14cm]{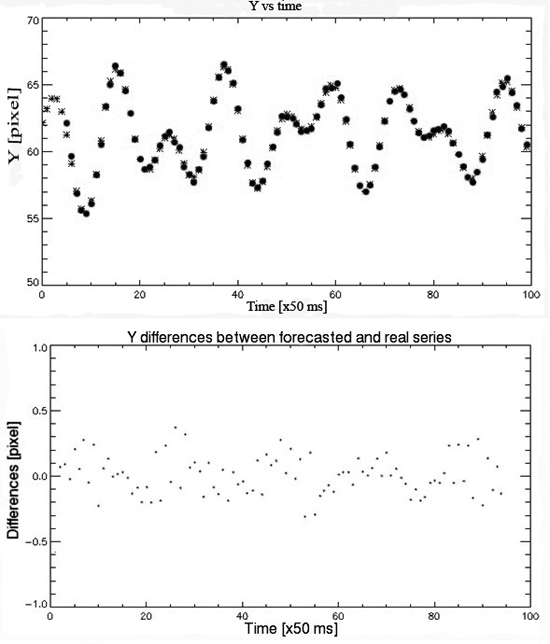}
\caption{\label{trajectory} {\it Upper Panel}: Y coordinate (of
the source centroid) corresponding to the 100 integrations
depicted in Figure~\ref{sum} ({\it Left Panel}): asterisks
correspond to the real (predicted) trajectory. {\it Lower Panel}:
Differences (in pixel) between the two trajectories. It is
worthwhile noticing the performance of the predictive algorithm
even working on rather complex trajectories and its promptness in
converging to the right prediction.}
\end{figure}

The final star image FWHM resulted just 10\% worse than that
corresponding to individual images (see Figure~\ref{sum}). The
plots in Figure~\ref{trajectory} depicts real and predicted
trajectories, together with their difference. Such difference
presents a standard deviation of 0.18 and 0.05 pixel in Y and X
direction, respectively. It is worthwhile noting that our
processing time is accomplished in real time (ie. it is much
shorter than the sampling time).

\vspace{3cm}

\vspace{3cm}

\subsection{Closed Loop Experimental Hardware for predictive fine tracking}

\begin{figure}[t!]\hspace{2cm}
\includegraphics[width=10cm]{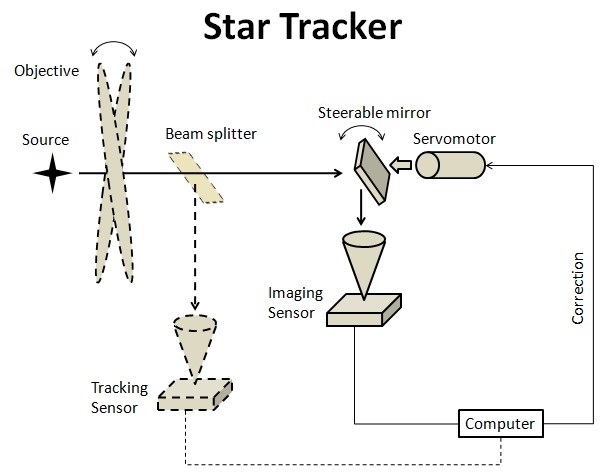}
\caption{\label{layout} Layout of our fine tracking system. Dashed components, although fundamental
to improve its performance, have not been used during the laboratory tests.}
\end{figure}

\begin{figure}[t!]
\includegraphics[width=14cm]{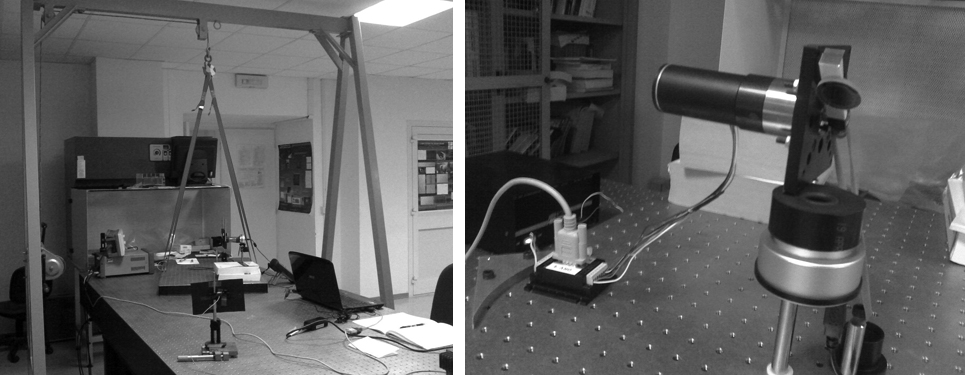}
\caption{\label{photos} {\it Left}: the realized device on our optical bench. {\it Right}: the correction system composed by
the CMOS camera, the movable mirror, and the servo-motor (by Faulhaber).}
\end{figure}

\begin{figure}[t!]
\includegraphics[width=14cm]{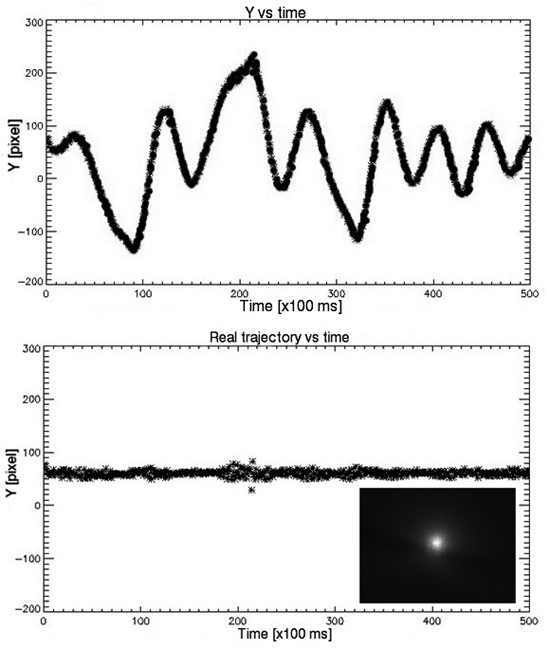}
\caption{\label{final} {\it Upper Panel}: asterisks indicate the Y
original trajectory (of the source centroid without any
correction; superimposed dots represent the predicted positions.
{\it Lower Panel}: focal plane trajectory corrected in real time
by both the steerable mirror and the predictive algorithm. In the
right bottom corner the final co-added image is shown.}
\end{figure}

The capability of our system to work in a closed loop
configuration has been tested by exploiting the experimental
layout shown in Figures~\ref{layout} and \ref{photos}. To verify
our predicting model, just one axis has been corrected by using
only one camera. Indeed, two cameras should be used, one dedicated
to the acquisition of the final image, and the other one as a
sensor for the tracking system. Instead, our simplified
configuration forced us to feed the predictive algorithm with the
{\it re-constructed} position of the star, and then to accept a
larger error induced by the backlash present in the steerable
mirror. In Figure~\ref{final} some test results are shown: while
the re-centering via software can be considered as an excellent
result (see Figures~\ref{sum} and \ref{trajectory}), the result
obtained with the movable mirror is not at the same level of
goodness, since the used mechanics is not precise enough. However,
we note (see Figure~\ref{final}) that oscillations larger than 300
pixels are reduced down to about 30 pixels, which represents the
limit imposed by the system backlash.

To get our final goal (to have error tracking less than 1 pixel)
we plan two future actions: ({\it i}) to improve the quality of
the steering mirror; ({\it ii}) to employ a two camera system
(imager and star tracking) so avoiding any systematic error. The
performances offered by the currently available tip-tilt optical
modules working at closed loop are more than enough for our scope.

\vspace{3cm}

\vspace{3cm}

\section{Concluding remarks and Perspectives}

The experimental tests of our predictive algorithm have
demonstrated it works very fine. A new tip-tilt system of higher
performance represents a next step of implementation. A rough
analysis of the S/N ratio of a stellar source at the focal plane
of a class 2m telescope, shows that a centroid of 0.1 arcsec is
easily reachable for a 16.5 mag star with an integration of just
0.03 sec. Hence, the fine tracking of the slow gondola
oscillations is possible on a wide sky area up to high galactic
latitudes, providing an available field of 50 arcmin$^2$, at
least. In fact, the model of Bahcall \& Soneira (1980) indicates a
density of 220 stars/deg$^2$ (brighter than 15.5 mag) at a
latitude of 90$^{\circ}$ (a sky areas poorly populated).
Therefore, in a field of 7$\times$7 arcmin$^2$, about 3 stars are
always present and suitable for the balloon fine tracking. At the
light of the presented considerations, the fine tracking topic
appears as a solved problem towards the flight of an optical-IR
telescope on board a stratospheric balloon.

\end{document}